\documentclass[aps,twocolumn,superscriptaddress,showpacs,preprintnumbers,amsmath,amssymb,floatfix]{revtex4-1}

\usepackage{graphicx,tabularx}
\usepackage{dcolumn}
\usepackage{bm}
\usepackage{color}

\usepackage{upgreek}

\renewcommand{\figurename}{Fig.~}
\newcommand{\equationname}{Eq.~}

\begin{document}


\title{Nonintrinsic origin of the   magnetic-field-induced metal-insulator and electronic
 phase transitions in graphite}

\author{Jos\'{e} Barzola-Quiquia} \email{j.barzola@physik.uni-leipzig.de}

\author{Pablo D. Esquinazi} \email{esquin@physik.uni-leipzig.de}

\author{Christian E. Precker}

\author{Markus Stiller}

\author{Mahsa Zoraghi}
 \altaffiliation[Current address: ]{Department
  of Neurophysics, Max Planck Institute for Human Cognitive and Brain
  Sciences, 04103 Leipzig, Germany}
\affiliation{Division of Superconductivity and Magnetism, Felix-Bloch-Institut f\"ur Festk\"orperphysik, Universit\"at Leipzig, Linn\'estrasse 5, D-04103 Leipzig, Germany}

\author{Tobias F\"orster}

\author{Thomas Herrmannsd\"orfer}
\affiliation{Hochfeld-Magnetlabor Dresden (HLD-EMFL), Helmholtz-Zentrum Dresden-Rossendorf, D-01328 Dresden, Germany}

\author{William A.  Coniglio}
\affiliation{National High Magnetic Field Laboratory, 1800 E. Paul Dirac Dr., Tallahassee, Florida 32310-3706, USA}

\date{\today}

\begin{abstract}
  A detailed magnetoresistance study of bulk and microflake samples of highly oriented pyrolytic graphite
  with a   thickness of 25~$\upmu$m  to 23~nm  reveals that the usually observed  field-induced metal-insulator
  and  electronic phase transitions vanish in thinner samples.
  The observed suppression is accompanied by orders of magnitude   decrease of the magnetoresistance
  and of the amplitude of the Shubnikov-de-Haas oscillations.
  The overall behavior  is related to the decrease in the quantity of
  two-dimensional interfaces  between crystalline regions of the same and different
  stacking orders present in  graphite samples.    Our results indicate that  these
field-induced transitions are not intrinsic to the ideal graphite structure and, therefore,   a relevant portion of
the published interpretations should be reconsidered.
\end{abstract}

\maketitle


In the early 1980's,  Tanuma~et al. reported a sharp increase in the
magnetoresistance (MR) of graphite when a strong magnetic field $\upmu_0 $H$ \ge 20~$T is applied
parallel to the \textit{c}-axis at temperatures $T < 10~$K~\cite{TANUMA-81}. Later, this
observation  was confirmed in different kinds of graphite
samples, e.g.~in Kish graphite~\cite{YAGUCHI-98,AKIBA-15,UJI-98,TIPM-83,FAUQUE-13,TAEN-18,tae18}, synthesized as a byproduct in steelmaking,
and in highly oriented pyrolytic graphite (HOPG)~\cite{IYE-82,IYE-84,KOPE-09}; for recent
reviews see \cite{YAGUCHI-09,chap4}.
There are a number of theoretical studies trying to provide an answer to several details of the
electronic transitions observed  in graphite at high fields and low temperatures.
Yoshioka and Fukuyama~\cite{YOSHIOKA-81}, for example,
proposed the  existence of
charge-density waves  or spin-density waves to explain such
electronic anomalies. Also, an excitonic
BCS-like state was proposed to understand the behavior observed
at fields above 50~T \cite{AKIBA-15}.  Recently,   the field-induced metal-insulator transition in thin flakes
of Kish graphite with thickness $t=178~$nm and 70~nm
 were studied~\cite{TAEN-18}, also under the influence of an electric field \cite{tae18}.
Those results were tentatively interpreted suggesting
that the electronic state in the insulating phase has an order along the stacking $c-$axis direction~\cite{tae18}.

Several unclear experimental details of the field dependence
of the electrical resistance of graphite samples added to the different
interpretations demonstrate that  there is no consent  on the
origin of
the  field-induced transitions.
Part of the reason is related not only on details of the
proposed phase diagram~\cite{YAGUCHI-09,chap4}
but to conflicting experimental evidence, as for example that
the electronic transitions are sometimes  absent
in certain ordered graphite samples~\cite{IYE-82,BRANDT-74}.
In this letter we show that these high-field transitions as well as
the metal-insulator transition are not intrinsic
of the graphite ideal structure but related to internal two-dimensional (2D) interfaces.

\begin{figure}
\includegraphics[width=\columnwidth]{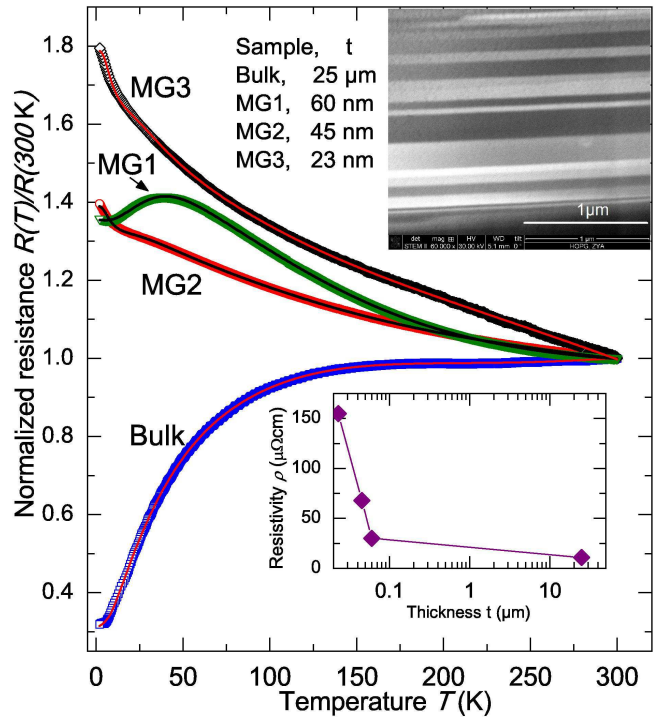}
\caption{\label{fig:fig1} Normalized resistance  of the
  investigated samples vs.~temperature.
The lines through the data points are fits to the three contributions in parallel as described in detail in Refs.~\cite{gar12,MZ-17,zor18}.
The main change between the bulk and the thinner flakes is given by the weight of the metalliclike
interfaces conductance.
The inset at the bottom shows the resistivity
  as function of the thickness $t$ at room temperature. The upper inset
  shows a STEM image with the e-beam parallel to the graphene planes of graphite. The
   $c$-axis of the graphite structure is normal to the interfaces existing between crystalline regions, shown with
different brightnesses. Those regions correspond either to crystalline Bernal regions twisted
 around the common \textit{c}-axis or to the
rhombohedral phase. For further STEM pictures see \cite{chap7}.}
\end{figure}

We start by pointing out a misleading assumption used  in the related literature, namely,  that the measured
electrical properties of graphite correspond to  the one of a homogeneous (structurally and electrically)
graphite sample~\cite{samples}.
Graphite is a layered material built by weakly coupled
graphene sheets, where usually the graphene layers adopt a hexagonal
$ABABA...$~(2H) stacking sequence (Bernal)~\cite{BERNAL-24}
  or as a minority phase the $ABCABCA...$ stacking order (rhombohedral (3R))\cite{LIPSON-42}.
 Scanning transmission electron microscopy (STEM)
measurements  show that
most of the samples are formed by a stacking of crystalline blocks with well defined
interfaces between them, see the upper right inset in Fig.~\ref{fig:fig1}, as example. In general, three types
of interfaces can be found, namely, between twisted 2H crystalline regions (we name it type I);
between twisted 3R regions (type II) and between (twisted) 3R/2H regions (type III).
The twist angle $\theta_{t}$  between the two crystalline regions of an interface is defined
through a rotation around the common $c-$axis (see, e.g., \cite{chap7}). It may play a main role in the electronic properties of a given interface. For example, Van Hove singularities in the density of states are situated closer
to the zero bias energy at  smaller $\theta_{t}$ \cite{bri12}, or a flat band is expected at $\theta_{t}=0^\circ$ for a type III interface \cite{mun13,kopbook}.
The thickness
of the crystalline regions (in the $c-$axis direction) having a common interface varies between $\sim 10$~nm to $\sim 500$~nm upon sample and location within the same sample. Electron back scattering diffraction (EBSD) indicates that the
lateral sizes of those crystalline regions in HOPG samples range between
$\sim 1~\upmu$m to $\sim 20~\upmu$m~\cite{GONZ-07}.

Earlier experimental studies  reported
the vanishing of the Shubnikov-de-Haas (SdH) oscillations amplitude the thinner the Kish graphite sample \cite{oha00,oha01}.
Furthermore, a nonlinear  increase of the resistance
of graphite samples by decreasing their thickness was reported ~\cite{Zhang-05}, i.e.,  the absolute resistivity increases
the thinner the sample \cite{JBQ-08,zor18}. All this experimental
evidence is at odd with the hypothesis of homogeneity assumed
in most of the
investigations of the electronic transport properties of graphite and speaks for
an unconventional contribution of  interfaces embedded within a semiconducting matrix.
We further note that superconductivity
at $\sim 1~$K  was discovered in  a single
interface, a twisted bilayer graphene \cite{cao18}. Additionally,
evidence for granular superconductivity with much higher critical temperatures at embedded
interfaces in bulk HOPG and natural graphite samples was reported earlier \cite{Ball-13,bal14I,bal15,pre16}.
Therefore, we expect that the main MR signal measured at low enough temperatures
and thick enough graphite samples
should be mainly related to the
 electronic systems  within the 2D interfaces.

We studied the in-plane MR  under pulsed
magnetic fields  $\upmu_0$H$ \leq 62$~T (applied parallel to the $c-$axis)  in
four different samples with thickness between $23~{\rm nm} \leq t \leq 25~\upmu$m
and lateral size from mm to below $10~\upmu$m, obtained from  a millimeter size HOPG
 sample from Advanced Ceramics (grade A). Further sample  details and measuring techniques
are given in the supplementary information (SI).  The MR measurements under pulsed fields
were accompanied by the temperature
dependence of the resistance $R(T)$ and MR
measurements under stationary
magnetic fields  to 18~T, shown in the SI.



Figure~\ref{fig:fig1} shows the temperature dependence of the resistance $R(T)$ of all four samples without applied field.
In the inset of Fig.~\ref{fig:fig1} the resistivity $\rho$  is plotted as
a function of sample thickness
$t$.
The temperature dependence of the electrical resistance
can be very well understood assuming the parallel contribution
of semiconducting regions with both stable stacking orders and a
metalliclike contribution from the interfaces~\cite{gar12,MZ-17,zor18}, as shown by
the fits to the data in Fig.~\ref{fig:fig1}.
Whereas the resistance of the thickest sample  shows the typical
metalliclike behavior of bulk graphite, the $R(T)$ of the  microflakes tends to
a semiconductinglike behavior the smaller the sample thickness.
The change from metalliclike
to semiconductinglike behavior, decreasing sample thickness, is due to the reduction of the number
of highly conducting 2D interfaces \cite{zor18}. Obviously, the (low) field-induced
metal-insulator transition does not occur in thin graphite samples.
The $R(T)$ curves
shown in Fig.~1, as well as those obtained in more
than 20 samples from different origins and measured in different laboratories,
can
be very well described between 2~K and 1100~K  with a parallel resistor model \cite{gar12,MZ-17,zor18}.
The difference in the fit parameters of the four samples shown in Fig.~1 is mainly in the total conductance  of the interfaces, decreasing the thinner the sample \cite{MZ-17,zor18}.


\begin{figure}
\includegraphics[width=\columnwidth]{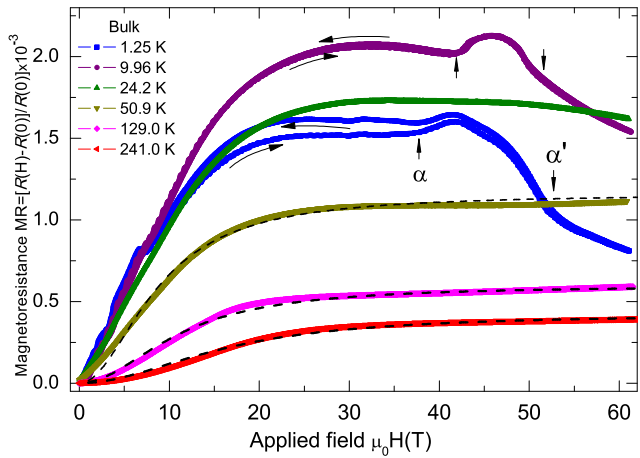}
\caption{\label{fig:fig2} High magnetic field results of the HOPG bulk
  sample ($t = 25~\upmu$m). The vertical arrows indicate the ``critical fields"
  $\alpha(T)$ and $\alpha'(T)$. The horizontal
  arrows  indicate the field sweep direction. The dashed lines through the data points at the three highest temperatures are fits to \equationname(\ref{eq:MR}) with the parameters  $\Delta n/n = 0.0484.$ and $\upmu = 1.277$~m$^2$/Vs at $T =241$~K. At lower $T$, $\upmu$ slightly increases whereas $\Delta n/n$ decreases, see SI for a discussion of the parameters. The best possible fit  of the data at 50.9~K to Eq.~(1) is
  shown only to emphasize the disagreement at low fields and the develop of a maximum at $\sim 30~$T.
  The fits to Eq.~(1) get much worse at lower $T$. }
\end{figure}

We now discuss the magnetoresistance, defined as MR$=(R($H$)-R(0))/R(0)$,
at different temperatures shown in Fig.~\ref{fig:fig2} for the bulk sample. In general, at $T \gtrsim 150~$K the contribution of the interfaces to the total MR starts to be overwhelmed by the higher conductance of the two semiconducting phases contributing in parallel~\cite{MZ-17}. Therefore,
at high enough temperatures the MR behaves as the one of a (low-gap) semiconductor. For graphite samples with lateral dimensions larger than the mean free path \cite{gon07,dus11,esq12}, the two-band model given by Eq.~(\ref{eq:MR}) and derived under the Boltzmann-Drude quasi-classical diffusive approach \cite{kelly}, provides a good (qualitative) description of the MR of bulk graphite at $T > 120~$K (dashed lines in Fig.~\ref{fig:fig2}).
The equation
\begin{equation}\label{eq:MR}
{\rm MR} = \left[\upmu^2 B^2 \left(1 - \frac{\Delta n^2}{n^2} \right)\right] /
\left [1 + \upmu^2 B^2 \frac{\Delta n^2}{n^2}\right],
\end{equation}
is a simplified version of the two-band
model equation assuming equal mobility for both electrons and holes
($ \upmu = \upmu _e \approx \upmu _h)$,
with $\Delta n/{n} = ({n_e - n_h})/({n_e + n_h}) $
 the relative charge imbalance between electron $n_e$ and hole $n_h$ carrier densities and $B = \upmu_0$H. This simplified
expression has only two adjustable fitting parameters, the
average mobility $\upmu$ and the relative charge imbalance
${\Delta n}/{n}$ and it is
insensitive to the absolute value of
$n_e$ (or $n_h$). Equation~(1) provides two key features of the experimental MR, namely the $B^2$ field dependence at low fields and its saturation
at high enough fields, see Fig.~\ref{fig:fig2}.

The MR data at $T \leq 50.9~$K shown in Fig.~\ref{fig:fig2} deviate from the predictions
of the two-band model (independently of the fitting parameters used in Eq.~(1)): there is a linear
field dependence at low fields and a maximum around 30~T develops.
The negative MR at high fields becomes more pronounced
the lower the temperature and, in addition,  a clear bump between $\sim 35$~T and  $\sim 55$~T  appears.
This behavior has been
reported for Kish and HOPG graphite and it was
attributed to field-induced phase transitions at  $T-$dependent critical fields $\alpha(T)$ and $\alpha'(T)$ (indicated by vertical arrows in Fig.~\ref{fig:fig2})
\cite{FAUQUE-13,IYE-82,YAGUCHI-09,UJI-98,chap4}. We further note that the overall MR decreases at $T < 10~$K. Also the absolute value of the resistance at high enough fields steadily decreases, i.e., $R(1.25$~K,~60~T) $\simeq 6~\Omega < R($241~K,~60T) $\simeq 9~\Omega$ (see Fig.S8 in the SI). This is attributed to  the so-called reentrance to a metallic state in the quantum limit, originally shown and discussed in \cite{yakovprl03}.

\begin{figure}
\includegraphics[width=\columnwidth]{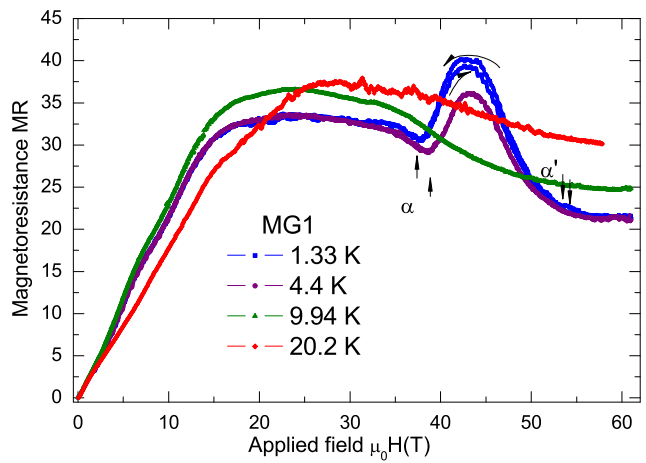}
\caption{\label{fig:fig3} High magnetic field magnetoresistance MR
  of sample MG1 ($t = 60~$nm)  at different temperatures. The vertical arrows
  indicate the critical fields $\alpha(T)$ and $\alpha'(T)$ and the horizontal
  arrows refer to the field sweep direction to emphasize the hysteresis observed only
  at the lowest temperature. At all other temperatures the curves are reversible in field within experimental resolution.}
\end{figure}

At low temperatures, a hysteresis  in the MR emerges at $\approx
12$~T and vanishes at $\approx 50$~T, being the MR smaller at the
increasing field branch, see Fig.~\ref{fig:fig2}. The opening of a
hysteresis at high fields was first mentioned by Takashi~et al.
\cite{TANUMA-81} and further discussed in~\cite{ARNOLD-17} using
data from a Tanzanian natural graphite sample. We stress that such
hysteresis in the MR is observed only at low temperatures and only
in thick enough samples (see also Fig.~\ref{fig:fig3}). See  the
SI  for more details on the mentioned hysteresis.

We discuss  the results of the thinner flakes. The MR results of  sample
MG1 ($t = 60$~nm) are plotted in~\figurename\ref{fig:fig3}.
One can clearly recognize the transitions at $\alpha(T)$ and $\alpha'(T)$.
  In general, the MR of this sample changes only
slightly in the measured temperature range  and behaves qualitatively
similar to the bulk sample, in spite of two to three orders of magnitude smaller sample width and length (see SI).
We note that the MR is $\sim 2$ orders
of magnitude smaller than in the bulk sample. In addition to the (small) reduction of the MR due to the decrease of the lateral size  (compared to
the bulk sample \cite{gon07})
the largest decrease of the MR is due to the decrease in the thickness and consequently in the
amount of interfaces~\cite{zor18}.
As in the bulk sample, a clear negative MR starts to appear at fields above $\sim 30$~T
and the field dependence is linear at low fields.

\begin{figure}
\includegraphics[width=\columnwidth]{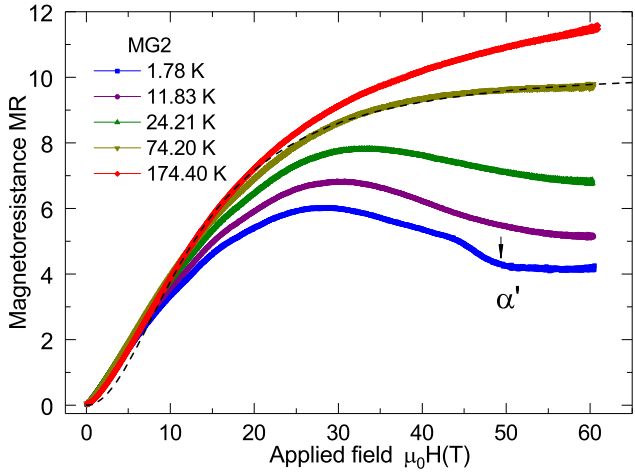}
\caption{\label{fig:fig4} High magnetic field magnetoresistance MR
  of sample MG2 ($t= 45$~nm) at different temperatures. The vertical arrow
  indicates the critical field $\alpha'(T=1.78$~K). The dashed line is
  calculated from Eq.~(1) to fit the high field MR data at 74.20~K, as example, although this
  equation is not really applicable due to non-diffusive, ballistic contribution, see text. Further, note
  the deviation of the data from the expected H$^2$ dependence at fields below 5~T.}
\end{figure}

Results of sample MG2 are plotted in Fig.~\ref{fig:fig4}.
In contrast to the previous samples, no evident field transition
$\alpha(T)$ is observed. At the lowest temperature we
can recognize a $\alpha'$~transition only. In the
 field range between 30~T and 62~T and increasing $T$ we  observe a change from
 a
negative to a positive MR. Note that the MR increases with $T$, without any sign of
saturation at high fields, in clear contrast to the bulk sample.
The MR is reversible within
experimental resolution.

\begin{figure}
\includegraphics[width=\columnwidth]{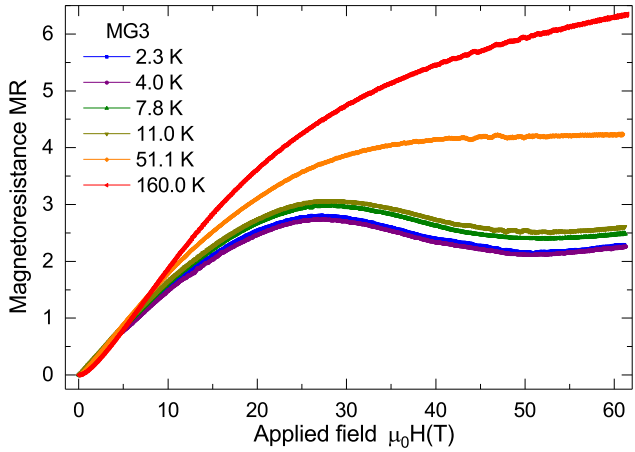}
\caption{\label{fig:fig5} High magnetic field  MR
  of sample MG3 ($t=23~$nm~$\sim  80$ graphene layers) at different temperatures.}
\end{figure}

\begin{figure}
\includegraphics[width=1\columnwidth]{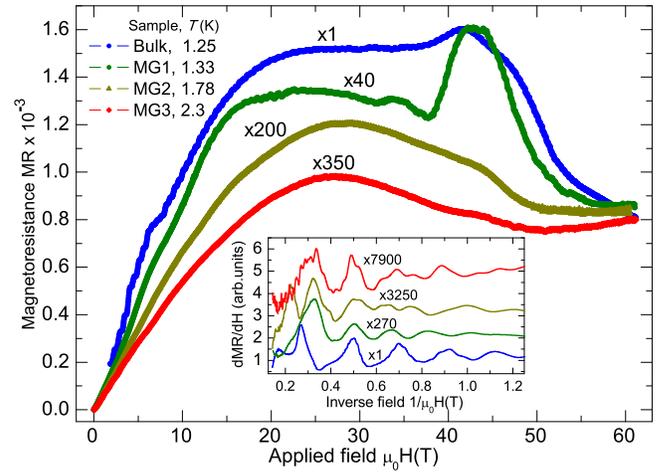}
\caption{\label{fig:fig7} Magnetoresistance of all samples investigated in this work
  at the lowest temperature. The inset shows
  the SdH oscillations obtained at $T=5$~K. The numbers beside the curves are the
  multiplication factor to show the results in the same scale. }
\end{figure}

The results of the thinnest sample MG3
are shown in \figurename\ref{fig:fig5}.
Its MR is overall much smaller than in the other samples and, as in sample MG2, its MR increases with temperature with no
sign of saturation. As in the other samples, at low enough $T$ and fields above 30~T, a negative MR  is observed to 50~T.
The MR of this sample shows only a smooth feature between 40~T and 50~T, i.e.,
where the high field-induced transitions were observed in the other samples, and at the lowest temperature.

In what follows we provide an interpretation of the main results. We note  that due to the different,
partially unknown, parallel contributions to the MR (one from the interfaces, which includes different types,  and two from the semiconducting layers \cite{gar12,MZ-17,zor18}),
the qualitative description we provide below uses the fact that the interfacial contribution to the total conductance overwhelms
the other two at $T < 100~$K \cite{MZ-17}. However, its influence on the MR weakens with fewer interfaces, as expected.\\
(1) The fact that the field-induced transitions at the fields  $\alpha(T), \alpha'(T)$ systematically vanish the smaller the thickness of the samples (for similar lateral sample dimensions) indicates that
these are
not intrinsic of the
ideal graphite structure. Taking into account previous galvanomagnetic studies \cite{JBQ-08,gar12,chap7,MZ-17,zor18}, their suppression is related to the smaller quantity of certain internal interfaces.
Our results and interpretation provide an answer to the absence of
high-field  electronic phase transitions in certain graphite
samples mentioned in Refs. \onlinecite{IYE-82,BRANDT-74}
as well as the scattering of the ``critical fields" data.\\
(2) The vanishing of the field-induced transitions is accompanied by a large decrease in the absolute MR in the
whole field range, see Fig.~6. The decrease in the MR by a factor  $\sim 700$ between the bulk and MG3  is mainly related to the decrease
in the amount of interfaces.\\
(3) The low-temperature  MR curves in Fig.~6 suggest that the field-induced transitions are superposed
with a MR curve that resembles that of the thinnest MG3 sample, i.e., the MR increases linearly in field at low fields, it reaches a maximum at 20~T$ < \upmu_0H < 30~$T  and shows a negative MR at  30~T$ < \upmu_0H < 50~$T (see also Fig.S9 in the SI). This fact added to the clear deviation from the expected two-band model behavior given by Eq.~(1) suggest that even the behavior of the thinnest sample MG3 at low temperatures is not yet intrinsic of the graphite ideal structure. Note that the SdH oscillations measured in graphite samples are not intrinsic but related to certain interfaces \cite{zor18}; these are observed in all samples, see inset in Fig.~6. However, the decrease of the amplitude of the SdH oscillations (characterized by the first field derivative of the MR) is more than 10 times larger than the decrease in the MR itself, suggesting that the MR of the graphite samples results from contributions of different interfaces or different regions within same interfaces. A considerable amount of extensive experimental work needs to be done to characterize the contribution(s) of each kind of interfaces to the total MR.\\
(4) Taking into account that certain interfaces can have granular superconducting properties in a broad temperature range \cite{Ball-13,bal14I,bal15,chap7,pre16,cao18},
 the linear MR at low fields as well as the negative MR at 30~T$~< \upmu_0H <50~$T may be
related  to the influence of the applied field to the Josephson
coupled superconducting regions. We note that  the normalized
resistance data of our samples resemble the behavior observed as a
function of field in granular low-$T_c$ superconductors, as AlGe
\cite{ger97} or InO \cite{gan96}, see discussion in Sec. II.A and
Fig.S9 in the SI. We refer the reader to the related literature \cite{bel07,var18}.\\
(5) Earlier experiments in thin graphite samples with no or a low number of interfaces showed that
the mean free path of the carriers within the graphene layers can be several microns large \cite{dus11},
of the order of our samples lateral size. In this case a ballistic, not the diffusive regime
assumed in Eq.~(1), should be taken into account to understand the non saturation of the MR
at high fields observed in the MG2 and MG3 samples (Figs.~\ref{fig:fig4}~and \ref{fig:fig5}).
The increase of the MR with temperature at all fields, as in the thinner samples MG2 and MG3, is observed because
the carriers mean free path, of the order of sample lateral size, decreases with temperature \cite{dus11,gon07}.\\
(6) Finally, we note that the MR oscillations periodic in field and the behavior under a bias voltage recently reported in thin graphite samples \cite{tae18} were already observed earlier and their origins are related to the existence of 2D interfaces and granular superconductivity \cite{esq08,bal14}.\\

Magnetoresistance measurements of graphite
samples of different thickness in a wide temperature and field range indicate that the
reported field-induced electronic phase transitions are not intrinsic of the ideal graphite structure but related to 2D electronic systems localized at certain interfaces formed between the
crystalline regions, commonly found in graphite samples with thickness above a few tens of nanometers.  Our conclusion is also supported by the thickness
dependence observed in other galvanomagnetic characterizations.
We encourage the
scientific community to revise the theoretical interpretations of the high-field transitions published in the past and to take into account explicitly the different kinds of possible interfaces graphite samples have.

Acknowledgements: We gratefully acknowledge A. Gerber (Tel Aviv University) and A. A. Varlamov
 (Istituto Superconducttori, Rome) for useful discussions on granular superconductivity and fluctuation effects, and P.K. Muduli for discussion on the two-band model. C.E.P. gratefully acknowledges the support provided by the Brazilian National Council for the Improvement of Higher Education (CAPES) under 99999.013188/2013-05.
The studies were supported by the DAAD Nr. 57207627
 (`Untersuchungen von Grenzfl\"achen in Graphit bei sehr hohen Feldern")
 and partially supported by the DFG under ES 86/29-1 and the SFB 762.
 A portion of this work was performed at the National High Magnetic Field Laboratory,
 which is supported by National Science Foundation Cooperative Agreement No. DMR- 1157490
 and the State of Florida. We acknowledge the support of the HLD at HZDR,
member of the European Magnetic Field Laboratory
(EMFL).



\newpage

\renewcommand{\figurename}{Fig.S\hspace{-3pt}}
\setcounter{figure}{0}

\begin{widetext}
\begin{center}
{\bf \large Supplementary information to: ``Nonintrinsic origin of the   magnetic-field-induced metal-insulator and electronic
  transitions in graphite"}
\end{center}

\end{widetext}
\vspace{2cm}

\section{Summary of samples dimensions and preparation details of the samples shown in the main manuscript}

The graphite microflakes were produced
by a rubbing method described in a previous
publication~\cite{JBQ-08}. After patterning the electrodes geometry for the resistance measurements
using electron beam lithography, the voltage and input current electrodes on the samples 
were produced by sputtering of Cr/Au.
Table I shows the dimensions of the four samples shown in the main manuscript. 
All samples were from the same bulk HOPG sample of grade A. 
All measurements were done in a four-probe configuration and magnetic
fields were applied along the $c$-axis direction of the graphite structure. 
The temperature dependence of
the resistance and the low field MR were initially characterized using
a commercial
$^4$He cryostat. The high magnetic fields magnetoresistance (MR) was measured at the high magnetic
field laboratory in Dresden (to $62$~T  applied with a
pulse length of $\sim 150$~ms) and in 
Tallahassee (DC fields to 18~T) within the temperature
range of $1.2$~K to $245$~K. A Lock-in amplifier (3.33~kHz) was
used to measure the voltage  during the rise and decay of the magnetic field. 
The applied currents varied between $5~\upmu$A to $10~\upmu$A
to avoid self heating effects.

In general, HOPG samples of  grade A 
thicker than $\gtrsim 100$~nm show transport properties
similar to bulk graphite. 
With exception of \cite{TAEN-18,tae18}, most of the previous studies on the high field MR of graphite
were done on thicker samples of millimeter
size~\cite{TANUMA-81,YAGUCHI-98,YOSHIOKA-81,AKIBA-15,IYE-82,KOPE-09,UJI-98,TIPM-83}.  
Taking into account the  internal structure of the graphite samples
\cite{gar12,GONZ-07,chap7} (see Fig.1 in the main manuscript), we need to reduce  the 
sample thickness to tens of nanometers and also the lateral size to a few
micrometers in order  to 
get electrical properties nearer to the intrinsic one of ideal, single phase graphite. 
\begin{table*}[]
\begin{tabularx}{\textwidth}{@{}XXXXXXXp{80pt}@{}}
\hline\hline
 Sample name & Length \textit{l}(m) & Width \textit{w}(m) & Thickness \textit{t}(m)   & $R(300$K$)~(\Omega)$   \\
 \hline 
  Bulk & 0.0035      &  $7\times 10^{-4}$        &      $2.5\times 10^{-5}$           &  0.023                     \\
	MG1      & $4\times 10^{-6}$        &  $7\times 10^{-6}$        &      $6\times 10^{-8}$             &  2.926                          \\
	MG2      & $4\times 10^{-6}$        &  $9\times 10^{-6}$        &      $4.5\times 10^{-8}$           &  6.7                       \\
	MG3      & $6\times 10^{-6}$        &  $8\times 10^{-6}$        &      $2.3\times 10^{-8}$           &  50.7                        \\
		\hline\hline
\end{tabularx}
\caption{Summary of samples dimensions and the absolute resistance at 300~K.}
\label{tab:Tab1}
\end{table*}

\begin{figure}
\includegraphics[width=\columnwidth]{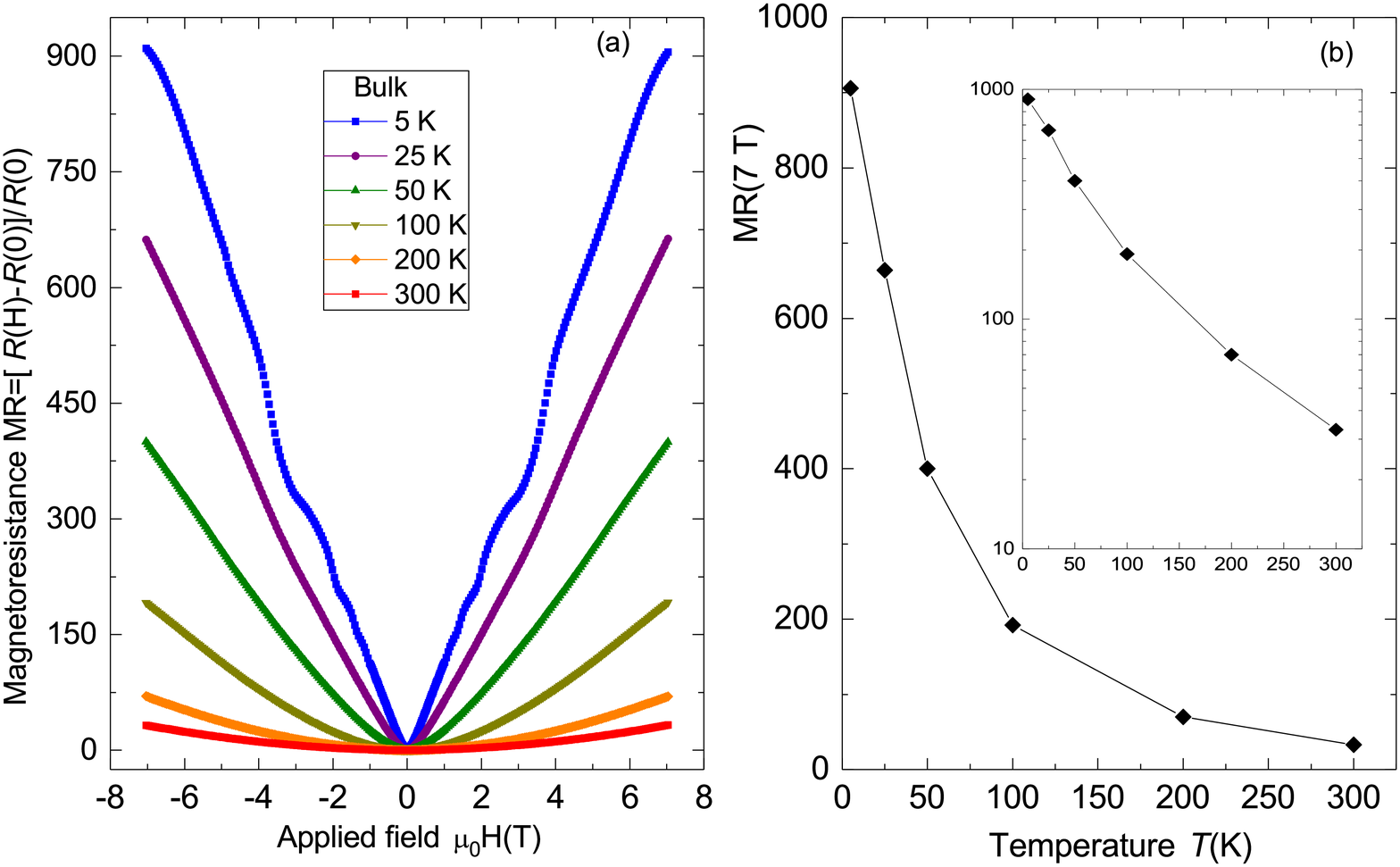}
\caption{\label{fig:fig1s} (a) Magnetoresistance of the bulk sample bulk measured at different constant temperatures.  (b) The temperature dependence of the MR at a fixed field of 7~T. Its inset shows the same data but in a semilogarithmic scale. }
\end{figure}
\begin{figure}
\includegraphics[width=\columnwidth]{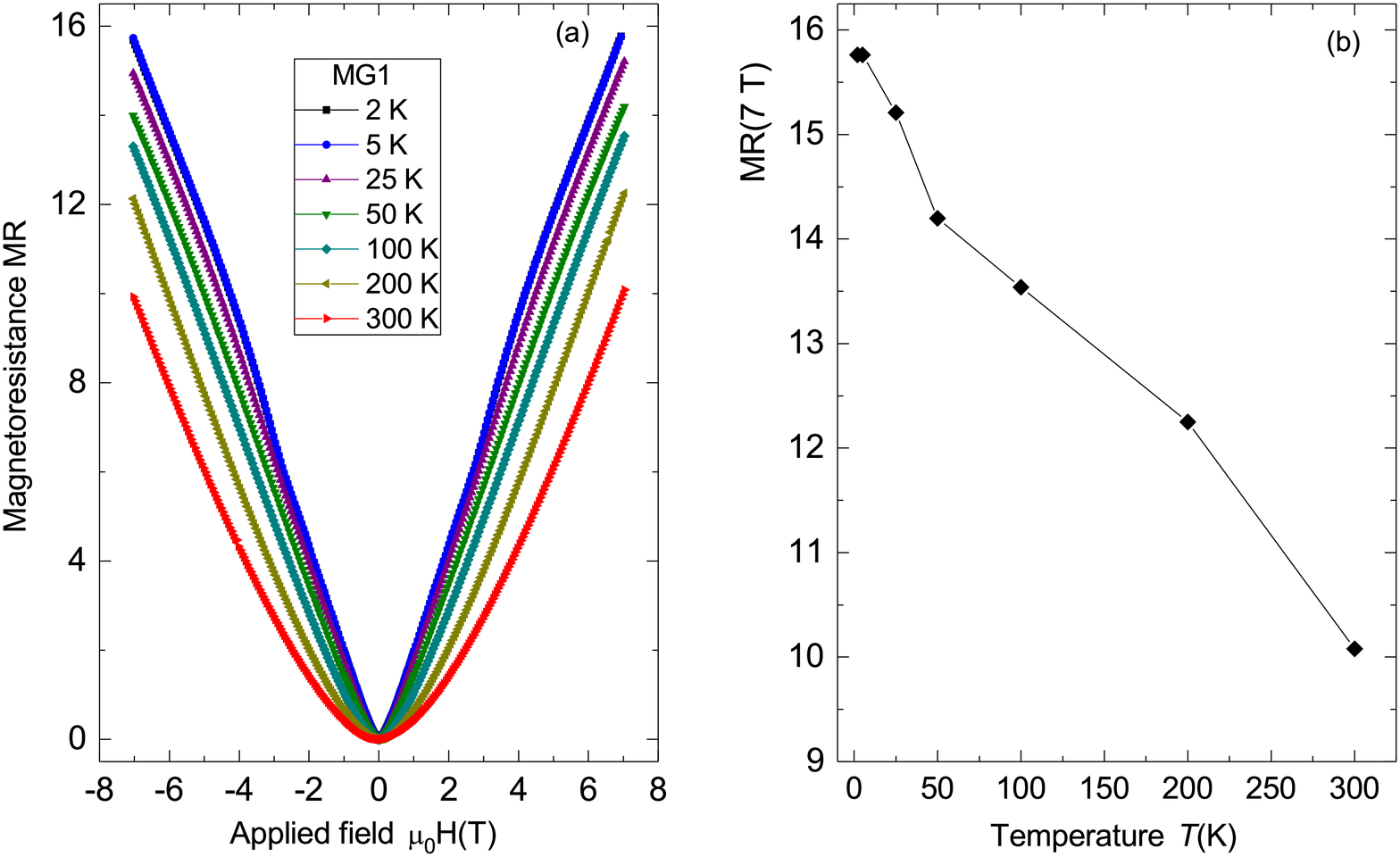}[b]
\caption{\label{fig:fig2s} (a) Magnetoresistance of the graphite flake MG1  measured at different constant temperatures. (b) The temperature dependence of the MR at a fixed field of 7~T.}
\end{figure}
\begin{figure}
\includegraphics[width=\columnwidth]{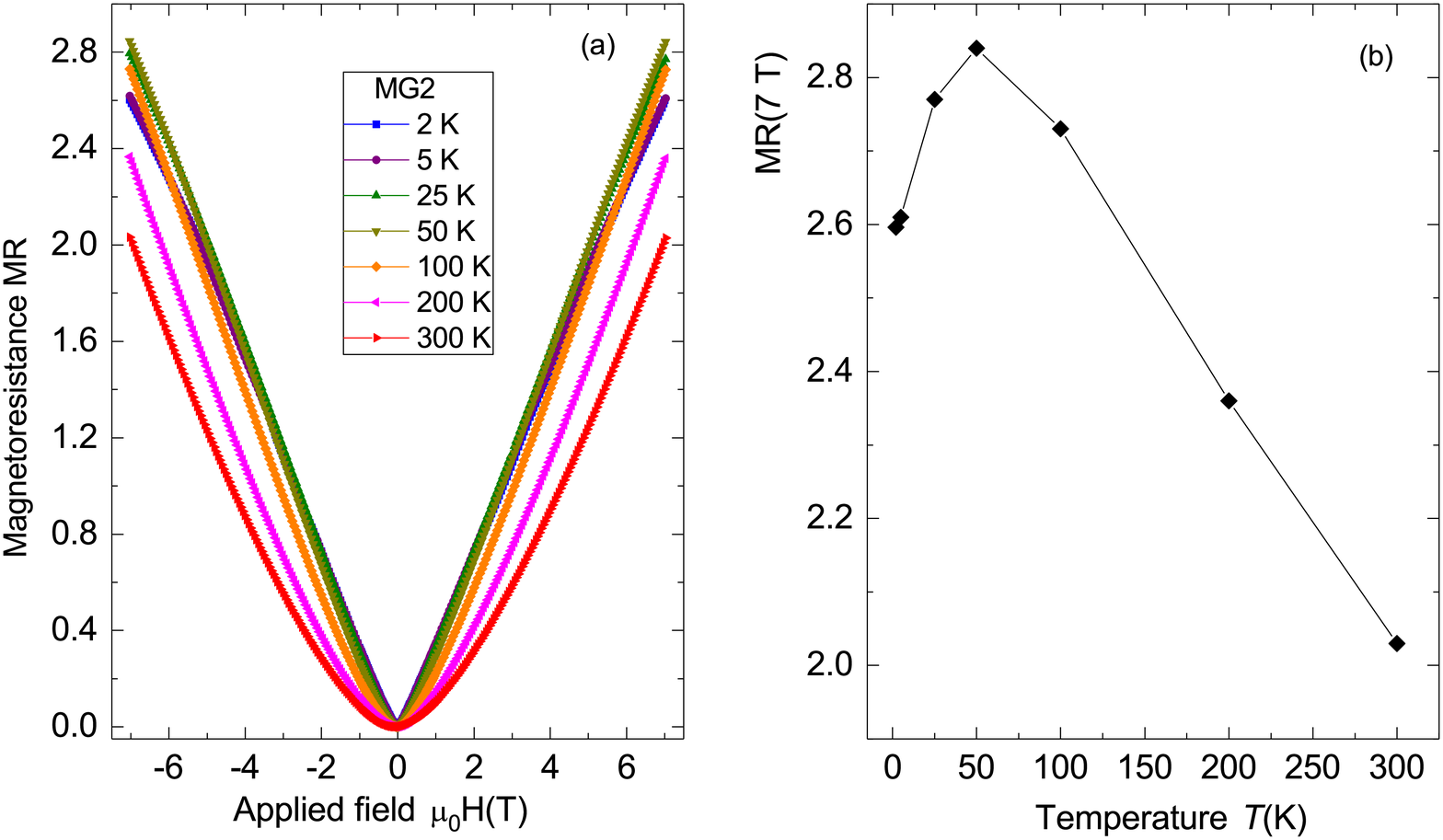}
\caption{\label{fig:fig3s} a) Magnetoresistance of the graphite flake MG2  measured at different constant temperatures. (b) The temperature dependence of the MR at a fixed field of 7~T. }
\end{figure}
\begin{figure}
\includegraphics[width=\columnwidth]{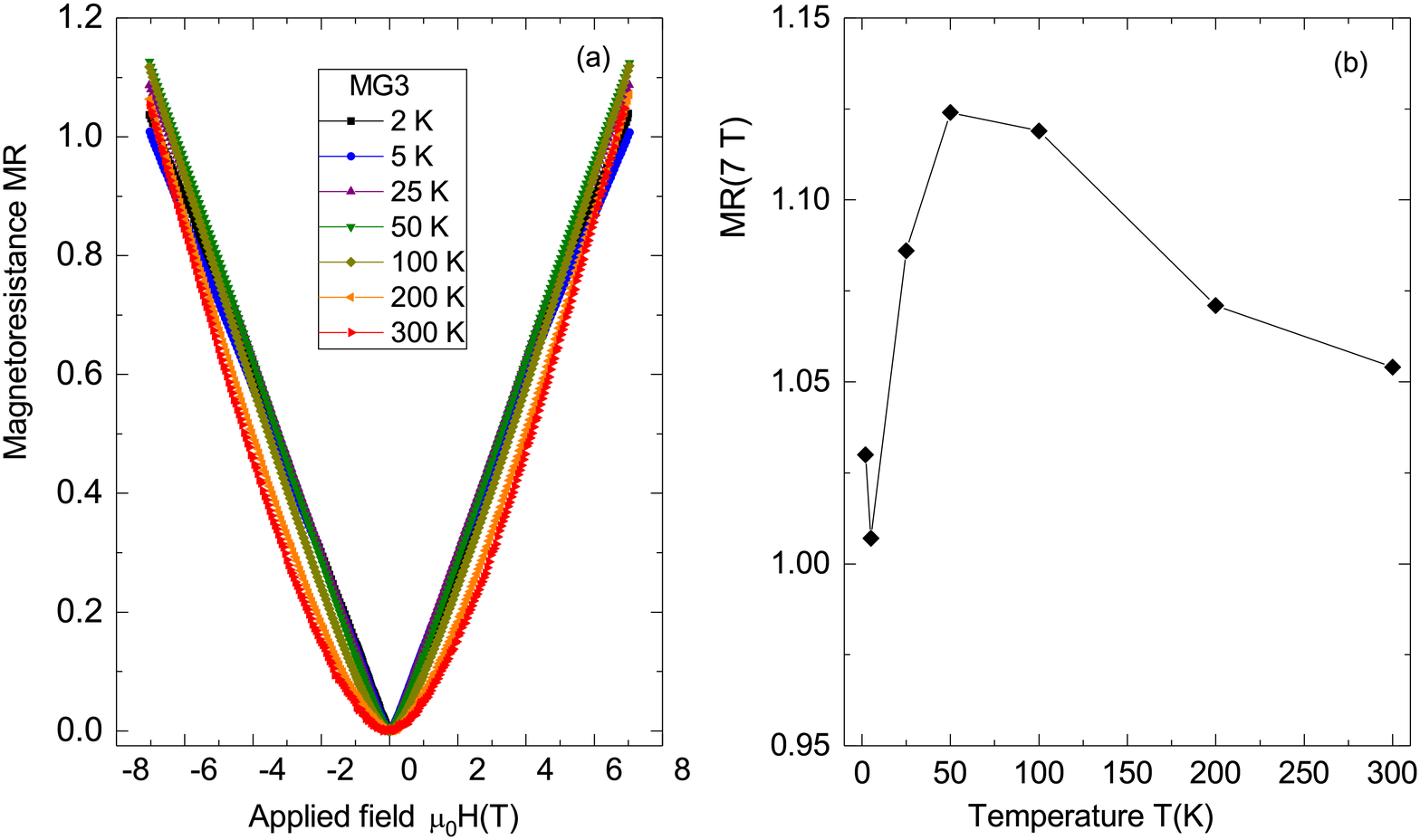}
\caption{\label{fig:fig4s} a) Magnetoresistance of the graphite flake MG3  measured at different constant temperatures. (b) The temperature dependence of the MR at a fixed field of 7~T.}
\end{figure}

\section{Results under DC applied fields}

\subsection{Temperature dependence of the magnetoresistance at fields  $\leq |\pm 7|$~T }

We first present the results at low fields  in the range $\pm 7$~T. The obtained results are plotted in~\figurename~\ref{fig:fig1s} to~\figurename~\ref{fig:fig4s}. In panels (a) we show the MR at different constant temperatures and in panels (b) the temperature dependence
of the MR at the fixed field of 7~T. The temperature dependence of the resistance at zero field is shown in Fig.1 of the main article. 
As pointed out in the main article, the MR systematically decreases the smaller the sample thickness. The main reason for this decrease is not
the change in the lateral size of the samples, which remains basically the same between samples MG1, MG2 and MG3, but the decrease in
the quantity of 2D interfaces. Interesting is also the temperature dependence of the MR at constant field shown in panels (b). 
Whereas for the thicker sample the MR at 7~T decreases by a factor of 30 between 5~K and 300~K, see Fig.~\ref{fig:fig1s}(b), this decrease with $T$
strongly diminishes for sample MG1 to a factor 1.6, see  Fig.~\ref{fig:fig2s}(b), and gets non-monotonous and overall much smaller for samples MG2 and MG3, see Fig.~\ref{fig:fig3s}(b) and Fig.~\ref{fig:fig4s}(b). Note that there is a relatively larger decrease of the MR with $T$ below 100~K for
samples bulk and MG1, i.e., in the region where the 2D interfaces contribution overwhelms the contribution of the semiconducting paths. 
The non-monotonous temperature behavior of MR($T$) below 100~K in the thinner samples MG2 and MG3 is related to 
the temperature dependent decrease of the carriers mean free path  $\ell(T)$  in the mainly semiconducting regions    and  the 
non-diffusive, ballistic transport that applies when 
$\ell(T)$ is of the order of the samples lateral size \cite{dus11,esq12}. The maximum in the MR$(T)$ shifts to higher temperatures the higher the applied magnetic field, see Figs. 4 and 5 in the main manuscript. 

To understand all the observed  effects we need to take into account: (a) the large carrier mean free path of
the carriers at the semiconducting graphene layers, which is comparable to the sample
size, (b)  the 
large Fermi wavelength $\lambda_F$ due to the low 
carrier density, (c) the decrease with  $T$ of $\lambda_F$ at the semiconducting regions, and (d)  the cyclotron radius $r_c$, which reduces with magnetic field \cite{gon07,dus11}. As discussed in [\onlinecite{gon07}] the MR vanishes when $\lambda_F \gg r_c$. 
We stress that the conduction mechanism in graphite samples occurs along the graphene layers and interfaces, there is no surface scattering.
The electron mean free path in thin graphite flakes is of the order of micrometers for  samples with thickness smaller than $\sim 50~$nm \cite{esq12,dus11}.
The transport in the $c-$axis direction, normal to the graphene planes, is negligible due to
the very weak coupling between the graphene planes in graphite. In fact, the MR depends only on the field component normal the graphene and interfaces planes. Deviations from this normal field component are due to intrinsic and/or extrinsic misalignments of the single crystallites within the 
graphite samples.

Note that the linear in field MR is observed at low enough temperatures and in all samples. 
Whether this linear in field dependence of the MR is intrinsic of the graphite structure or related 
to granular superconductivity is not yet clear, see the discussion in Section V below. 
We stress that even in the thinnest sample the contribution of the interfaces is still measurable 
at low enough temperatures, as the overall field dependence of the MR (with a maximum at a certain high field) and the existence of the SdH oscillations indicate, see Fig.6 in the main manuscript and its inset. We would like to note that the linear in field magnetoresistance is a subject discussed
in a large number of publications. The linear in field dependence of the longitudinal resistance is 
very often invoked as evidence for exotic quasiparticles in new materials. On the other hand linear magnetoresistance 
has been measured in ``simple" semiconducting samples like  Mn implanted Ge \cite{sim14} or in  2D electron gas
  in an ultrahigh mobility GaAs quantum well \cite{kho16}. 
  Experimental evidence suggests that its origin can be  an admixture of a component of the Hall resistivity 
  to the longitudinal resistance related to density fluctuations, which exist in nearly every sample, especially when the
  carrier density is low \cite{kho16}.

\subsection{The temperature dependence of the resistance and MR at fields $\leq |\pm~18|$~T}
\begin{table*}[]
\begin{tabularx}{\textwidth}{@{}XXXXXXXp{80pt}@{}}
\hline\hline
 Sample name & Length \textit{l}(m) & Width \textit{w}(m) & Thickness \textit{t}(m)   & $R(100$K)$~(\Omega)$   \\
 \hline 
  BSR (bulk) & 0.0032      &  $1.03\times 10^{-3}$        &      $5\times 10^{-5}$          &  0.00515                     \\
	MG-P1      & $14.5 \times 10{-6}$        &  $8.3 \times 10^{-6}$        &      $10 \times 10^{-8}$             &  5.69                                               \\
		\hline\hline
\end{tabularx}
\caption{Summary of the dimensions of two more graphite samples, which MR was measured to $\pm~18$~T. 
The BSR sample was  a natural graphite and sample MG-P1 was prepared from
a HOPG, grade A bulk sample.}
\label{tab:Tab1}
\end{table*}
\begin{figure}
\includegraphics[width=\columnwidth]{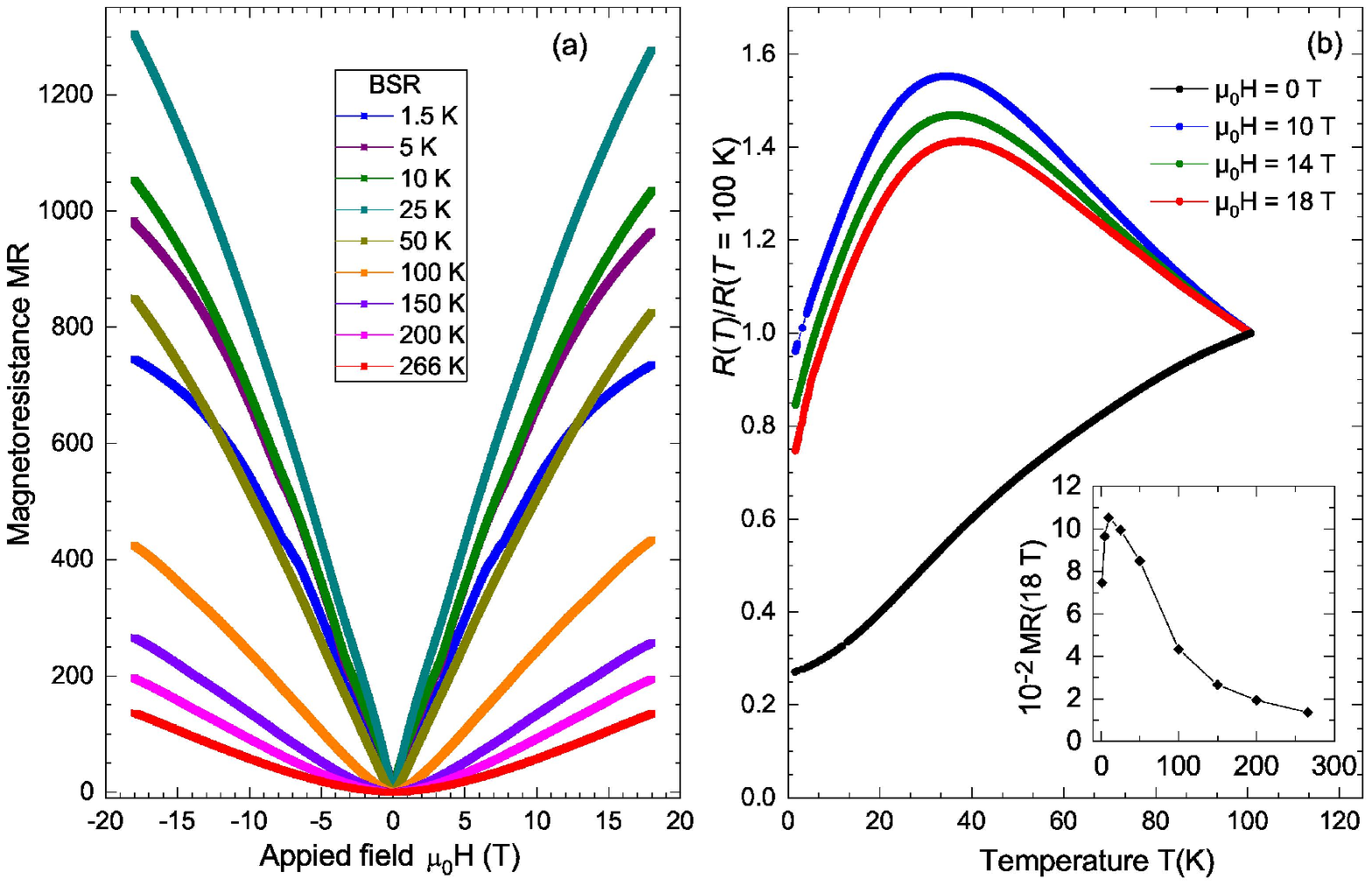}
\caption{\label{fig:fig6s} (a) Magnetoresistance and in (b), the temperature dependence of the resistance $R(T)$ at different constant applied fields of 
a second bulk graphite sample, see Table II. The inset shows the MR(18~T)~vs. temperature. }
\end{figure} 
\begin{figure}
\includegraphics[width=\columnwidth]{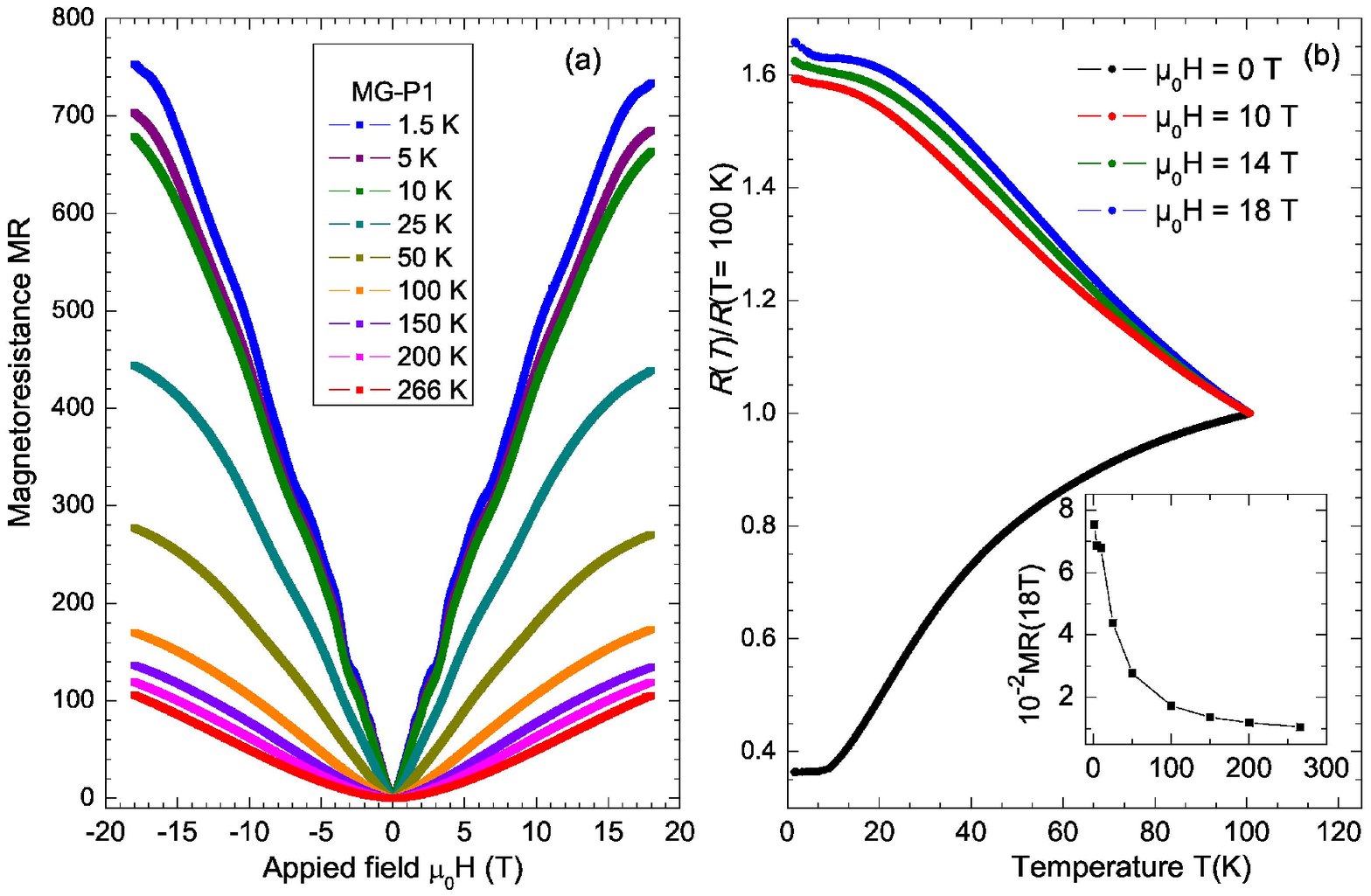}
\caption{\label{fig:fig7s} (a) Magnetoresistance and in (b), the temperature dependence of the resistance $R(T)$ at different constant applied fields of the
graphite flake (100~nm thick)  MG-P1, see Table II. The inset shows the MR(18~T)~vs. temperature. }
\end{figure} 
\begin{figure}
\includegraphics[width=\columnwidth]{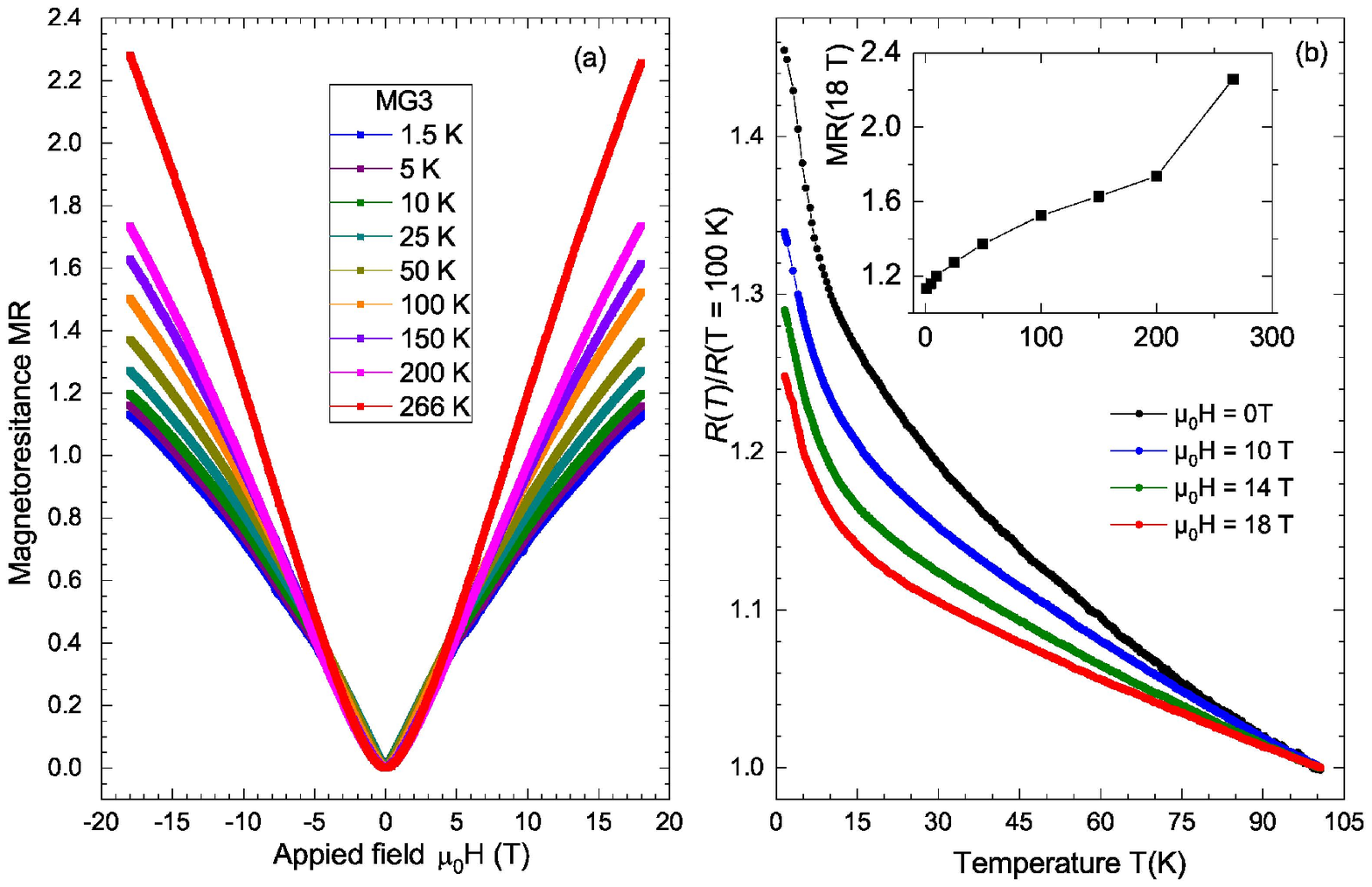}
\caption{\label{fig:fig8s} (a) Magnetoresistance and in (b), the temperature dependence of the resistance $R(T)$ at different constant applied fields of
the same flake MG3 measured with pulsed fields. Both sets of data are similar. The inset shows the MR(18~T)~vs. temperature. }
\end{figure}
Table II shows the dimensions of two more samples measured to $\pm~18$~T.  
 The temperature dependence of the resistance $R(T)$ at different constant fields as well as the magnetoresistance MR results are plotted in \figurename~\ref{fig:fig6s} to \figurename~\ref{fig:fig8s}. Similar to the bulk HOPG sample shown in the main manuscript and above, 
 the bulk BSR sample shows 
 a huge MR, which diminishes increasing the temperature. In   the inset of 
\figurename~\ref{fig:fig6s}(b) we have plotted the MR(18~T) as a function of  temperature. Its behavior is similar to other thick HOPG or Kish graphite samples.
The $R(T)$ results under applied fields of the bulk sample BSR shows the already reported reentrance to a metalliclike state at high enough fields and low enough temperatures, observed in HOPG bulk graphite samples\cite{KOPE-03}. 
 We would like to
remark that such reentrance   behavior is also thickness dependent. It is  clear to see the metalliclike behavior up to 30~K in the bulk sample (see \figurename~\ref{fig:fig6s} (b)) at  the applied fields.  While in the case of the thicker flake (see \figurename~\ref{fig:fig7s}(b)) a smooth but evident metalliclike behavior develops at $T < 20$~K. In  the thinnest flake MG3 and at all investigated fields no reentrance behavior is observed (see \figurename~\ref{fig:fig8s}(b)). All this evidence points out that the origin for the reentrance is related to the magnetic response of certain interfaces existent in thick enough samples.

\begin{figure}
\includegraphics[width=\columnwidth]{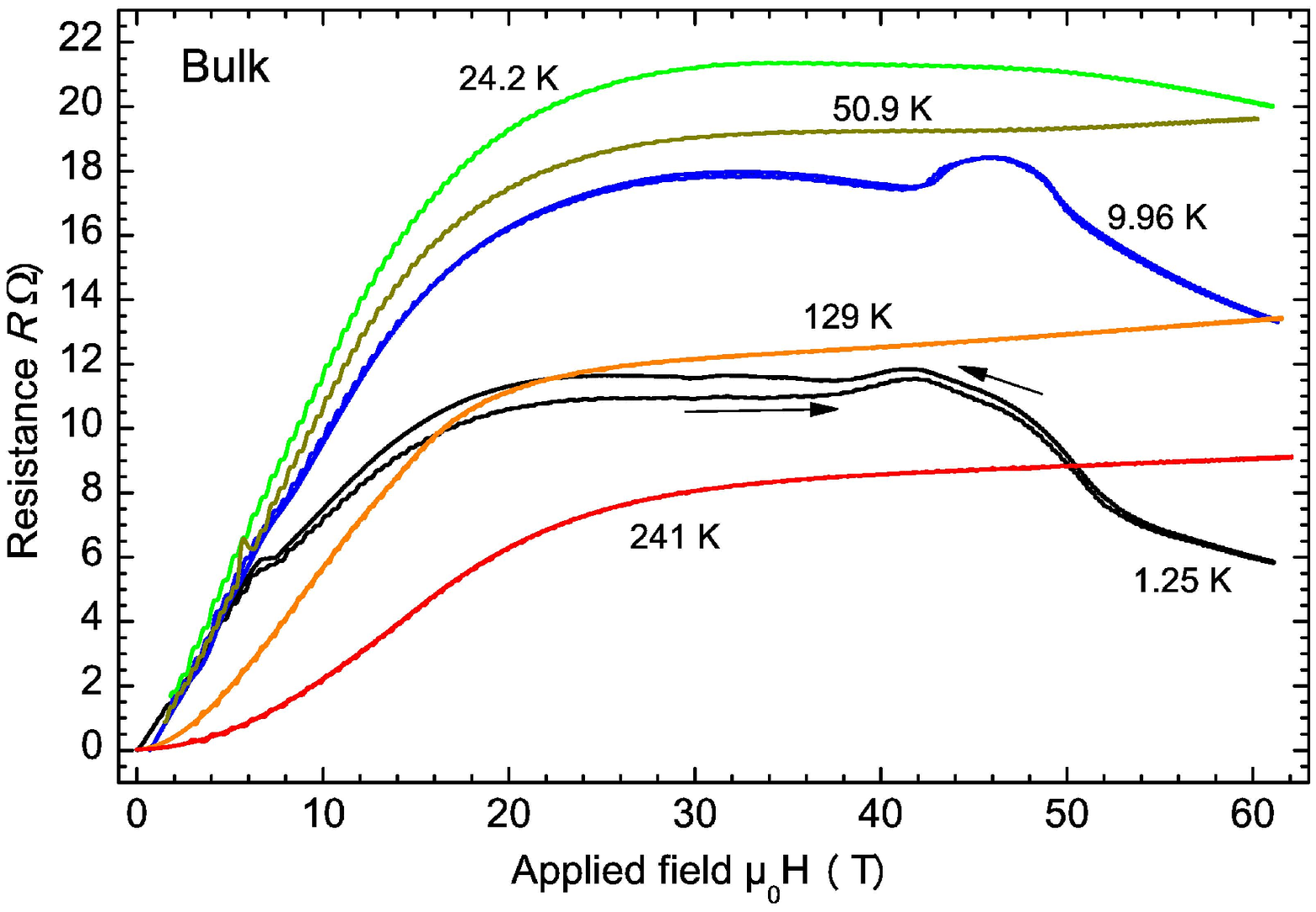}
\caption{\label{fig:fig9s}  Absolute resistance vs. magnetic field at different constant temperatures of the same bulk sample shown in the main
manuscript.  }
\end{figure}

\section{The reentrance to a metalliclike state at high magnetic fields in thick graphite samples}

The reentrance to a metalliclike state shown in  \figurename~\ref{fig:fig6s}(b) has been shown originally in Ref.~\onlinecite{KOPE-03} and
a possible explanation was given in terms of the enhancement of the density of states by the applied field and the possibility that
 superconducting correlations appear \cite{KOPE-03,yakovadv03}. Note that at the time of those publications the contribution of the embedded interfaces 
to the conductance of  graphite samples was not known and therefore it was not considered in the proposed explanations. 
 We would like to point out here that the observed reentrance provides a further evidence for the different contributions in parallel, i.e., for the
 inhomogeneous nature of thick graphite samples. To understand this point,  the best is to plot the data of Fig.~2 of the main manuscript in absolute
 resistance values, instead of the MR, as shown in  \figurename~\ref{fig:fig9s}. In this figure we note that the absolute resistance at 1.25~K and at fields above 50~T gets {\em smaller} than the resistance at 241~K at the same fields. This fact is very difficult to understand were the sample homogeneous, 
 independently of the reason for the negative resistance that develops above 30~T and at low enough temperatures. 
 However, if we consider that at 241~K the measured resistance is basically due to
 the semiconducting regions and at 1.25~K those semiconducting regions are effectively short circuited by the interfaces, the observed behavior is
much  easier to be understood, assuming that the maximum and the negative MR is a phenomenon that exists only at certain high 
conducting interfaces. A short discussion on the negative MR observed at high enough fields and low enough temperatures is given in Section~\ref{granular}.

\section{Parameters of the two-band model and the field dependence of the MR}

The field dependence of the resistance measured at high enough temperatures, as shown in 
\figurename~\ref{fig:fig9s}, see also Fig.~2 in the main manuscript, can be well explained with the
equation derived from the two-band model (Eq.~(1) in the main manuscript). The values of the fit parameters
should be taken, however, with some care. The mobility values we obtain between 50~K and and 241~K to fit
the MR field dependence shown in Fig.S8 are 3.95, 2.22 and 1.28 m$^2$/Vs. At lower temperatures the theoretical
curves strongly deviate from the experimental data, which show one or two maxima in the MR at certain fields. 
The obtained fit values of the mobility are in agreement with published values for very thin  \cite{zha05} and thick \cite{not75} 
graphite samples. On the other hand, direct measurements of the mobility of the carriers inside the semiconducting
graphene planes in thin graphite samples using micro-constrictions, provide values at least one order of magnitude larger
than those obtained from the fits at similar temperatures \cite{dus11}. The question whether this difference is because: (a) in thick
samples the MR field dependence is mainly given by the response of the interfaces, which have a much larger carrier density 
($n \sim 10^{10} \ldots 10^{11}~$cm$^{-2}$) than in the graphene planes of the semiconducting regions ($n \lesssim 10^{8}~$cm$^{-2}$), and/or (b) the
mobility values obtained from the two-band model are incorrect because the used Eq.~(1) (see main manuscript) is not, rigorously speaking, applicable when at least part of the carrier dynamic is not diffusive but ballistic \cite{gar08}, is not yet clarified. 

\section{The Hysteresis in the MR at high magnetic field observed in thick graphite samples}

The first reference for an absence of any field hysteresis in the MR between the up and down field sweeps is found in the paper of  
Iye~et al.~\cite{IYE-82}. The authors  mentioned there that because of this  the magnetic field induced transitions
$\alpha$ and $\alpha'$  were phase transitions of second or higher order.  In the publication of Miura~et al~\cite{MIURA-92}, high field MR measurements were done on Kish graphite. In their  Fig.~1 the MR measured at 4.2~K  clearly shows a field hysteresis, which begins at $\upmu_0$H$\approx 12$~T and ends at $\upmu_0$H$\approx 45$~T. However,  in the manuscript this hysteresis was not mentioned and discussed. Another work published by Ochimizu~et al~\cite{OCHIMIZU-92} also showed a field hysteresis in the MR  at different temperatures (see Fig. 2 in that publication) in a field range similar to that of Ref.~\onlinecite{IYE-82} and ours, see Fig.~2  in our main article. 
Recently, Arnold~et al. \cite{ARNOLD-17}  reported the MR of bulk graphite at high magnetic fields and found  sharp new features at  $\upmu_0$H$_{\alpha'}=52.3 \pm 0.1$~T and $\upmu_0$H$_{\beta'}= 54.2 \pm 0.1$~T. The authors suggested that they  correspond to two distinct
first order transitions
associated with the abrupt depopulation of both the $(0,\uparrow)$  (electron like) and $(-1,\downarrow)$ (hole like) Landau levels.
They suggested that the $\alpha'$ and $\beta'$ are first-order transitions on the basis of 
the observed field hysteresis, which is shown in  Fig.4(c) of that work. Independently of the theoretical explanation, our experimental results 
show  that such hysteresis  has a thickness dependence. The high fields hysteresis between up and down field sweeps is only observed at low enough temperatures and in  bulk and 
thicker samples ($t \ge 60~$nm).  

A general remark on studies under pulsed field. It is clear that  heating should be always considered, i.e., whether $\dot{{\rm H}}$ might affect the result. Therefore, it is necessary to provide the following data to
assure that the hysteresis is not an artefact: (1) One needs pulsed and DC field data at the same temperature. In our case would mean to
apply DC fields up to 30~T, at least, which is much above the maximum available;  (2) High field and low field pulses at the same temperature; (3) 
Fast and slow pulses (with up/down sweeps) at the same peak field and temperature.

\section{Granular superconductivity}
\label{granular}
\begin{figure}
\includegraphics[width=1.\columnwidth]{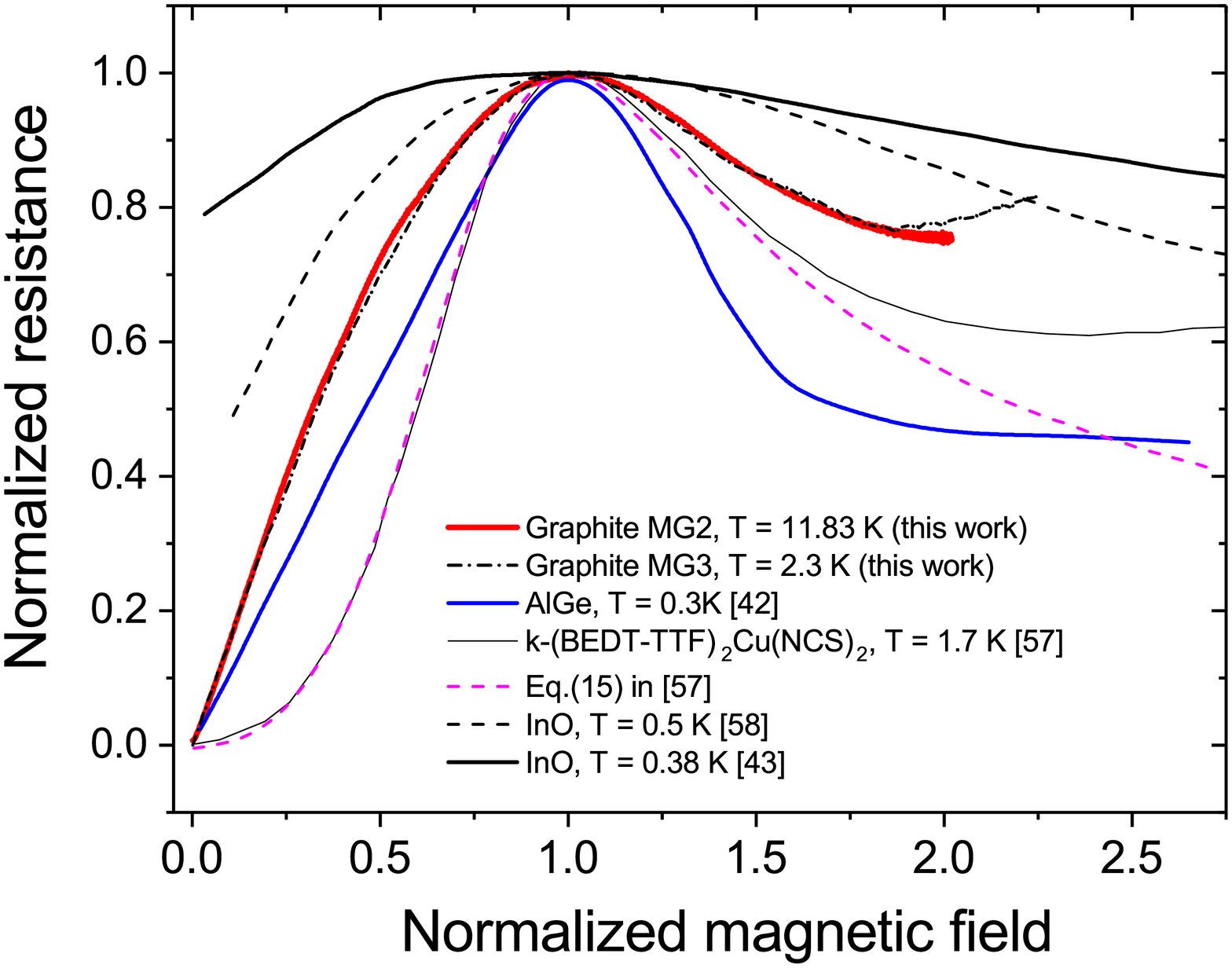}
\caption{\label{gra}  Normalized resistance vs. normalized magnetic field at different constant temperatures of: graphite MG2 and MG3 samples measured in this work, with the normalization field factor at resistance maximum of $\upmu_0$H$^\star = 27.5$~T and 30~T at 11.83~K and 2.3~K, respectively; the granular superconductor AlGe from Ref.\onlinecite{ger97} with $\upmu_0$H$^\star \simeq 2.3$~T and at 0.3~K; the organic layered superconductor $\kappa$-(BEDT-TTF)$_2$Cu(NCS)$_2$  from Ref.\onlinecite{gla11} with $\upmu_0$H$^\star \simeq 2.8$~T at 1.7~K with the voltage measured parallel to the applied field; a granular superconducting  InO thin film from Ref.\onlinecite{lee13} with $\upmu_0$H$^\star \simeq 2.4$~T at 0.5~K; InO thin film from Ref. \onlinecite{gan96} with $\upmu_0$H$^\star \simeq 3.0$~T at 0.38~K.}
\end{figure}

A key feature of the high magnetic field results in graphite samples is the maximum MR and the further negative MR observed above that maximum. As pointed out in the main manuscript and taking the normalized data of Fig.6 into account, we assume that this  feature is due to the magnetic field response of certain interfaces in the graphite sample and it is not
an intrinsic property of the ideal graphite stacking orders. Previous studies indicate the existence of granular superconductivity at certain interfaces with a rather broad range of critical temperatures upon sample, see  Refs.~\onlinecite{Ball-13,bal14I,bal15,chap7,pre16,sti18,cao18}. Therefore, we compare below our MR data with that obtained in granular superconductors at temperatures and fields below the critical values and in one case above the critical field. 

In Fig.S9 we show the normalized       MR data of the thinnest samples MG2 and MG3 at 11.83~K and 2.3~K and include the MR data of granular Al in a Ge matrix from Ref.~\onlinecite{ger97}, obtained at 0.3~K. It is  remarkable that the normalized MR data of our MG2 and MG3 samples are practically identical, pointing to a common origin. 
The MR data of granular Al/Ge show a linear field dependence at low fields and a clear negative MR in a field range comparable (in normalized units) to that of the graphite samples. In the case of granular Al/Ge the field at which the negative MR regions ends is considered as the upper critical field of the superconducting grains at $\upmu_0$H$_{c2}(0.3$K)$\simeq 4.5~$T. 
In case of the granular InO thin films \cite{lee13,gan96} a similar behavior is observed although the negative MR field range has a larger extend above the maximum at $\upmu_0$H$^\star$. The interpretation of the MR maximum, with a value larger than the resistance in the normal state at the same temperature, and the negative MR is given  in terms of Josephson coupled granular superconducting grains \cite{ger97}. As pointed out in Ref.~\onlinecite{gan96}, the negative MR at high fields can be described by a field dependent energy gap in the density of states at the Fermi level, which results from Cooper interactions. Furthermore, it has been argued that superconducting clusters at high enough fields can appear due to fluctuations \cite{gan96,var18}, a phenomenon that may play a role in 2D superconducting layers. 

The effect of superconducting fluctuations above the critical temperature and field have been 
studied in detail in the last years, see Ref.~ \onlinecite{var18} and Refs. therein. As example,     
we include in Fig.~S\ref{gra} the MR data of M. Kartsovnik obtained from the 
layered organic superconductor $\kappa$-(BEDT-TTF)$_2$Cu(NCS)$_2$ at 1.7~K with the theoretical line given in Ref.~\onlinecite{gla11}. One main difference with respect to the other data shown in that figure is that the MR follows a quadratic instead of a linear 
field dependence at low fields. Also, the theory \cite{gla11} is applicable in the normal state of a granular superconductor. From all the 
evidence obtained during the last years in graphite
we believe that the temperature where the maximum is observed are well below the critical temperature. Assuming that the field at which the negative MR ends (in AlGe that would be $\sim 1.8$ times the field at the maximum MR) is related to the critical field, i.e., $\upmu_0$H$_{c2}(0) \sim 60~$T for the granular superconducting 2D interfaces in graphite.


\begin{thebibliography}{59}
\bibitem [{sam()}]{samples}%
  \BibitemOpen
  \href@noop {} {}\bibinfo {note} {{\rm The electrical and structural
  homogeneity was assumed in the literature independently whether the sample
  was natural graphite, highly oriented pyrolytic graphite (HOPG) or Kish
  graphite.}}\BibitemShut {Stop}%
\end{thebibliography}

%

\end{document}